\title{\textbf{Generalized beta convolution model of the true intensity for the Illumina BeadArrays}}
\author{
        \large
        \tt {Rohmatul Fajriyah} \\
        \small
	\tt{Institute of Statistics, TU Graz, Austria} \\
        \small
        \tt{Dept. of Statistics, Univ Islam Indonesia, Jogjakarta, Indonesia} \\
	\href{mailto:fajriyah@student.tugraz.at}{fajriyah@student.tugraz.at}, \href{mailto:rfajriyah@fmipa.uii.ac.id}{rfajriyah@fmipa.uii.ac.id}
}
\date{}

\documentclass[12pt]{article}
\usepackage[left=1.5in, right=1in, top=0.75in, bottom=0.75in, includefoot, headheight=13.6pt]{geometry}
\usepackage[colorlinks=blue]{hyperref}
\usepackage{commath,subfig,graphicx,amssymb,amsthm,amstext,amsmath,mathtools,mdwlist,color,float,rotating,rotfloat,caption,xtab,setspace,relsize,exscale,titlesec,soul}
\newtheorem{definition}{Definition}[section]

\usepackage[square, comma, numbers, sort&compress]{natbib}
\usepackage[T1]{fontenc}
\setulcolor{blue}
\setstcolor{red}
\sethlcolor{green}
\numberwithin{figure}{section}
\usepackage{sectsty}
\newtheorem{theorem}{Theorem}[section]

\begin{document}
\maketitle

\begin{abstract}
Microarray data come from many steps of production and have been known to contain noise. The pre-processing is implemented to reduce the noise, where the background is corrected. Prior to further analysis, many Illumina BeadArrays users had applied the convolution model, a model which had been adapted from when it was first developed on the Affymetrix platform, to adjust the intensity value: corrected background intensity value.

Several models based on different underlying distributions and or parameters estimation methods have been proposed and applied. For instance : the exponential-gamma, the normal-gamma and the exponential-normal convolutions with a maximum likelihood estimation, non-parametric, Bayesian and moment methods of the parameters estimation, including two recent exponential-lognormal and gamma-lognormal convolutions.

%However none of these existing models, in the benchmarking study, performs outstandingly to others. Each of them has its own drawback. Therefore, the need to build the new model is still widely opened.
In this paper, we propose models and derive the corrected background intensity based on the generalized betas and the generalized beta-normal convolutions as a generalization of the existing models.
\end{abstract}

\vskip 3mm

\newpage{}
\pagenumbering{arabic}
\section{Introduction} \label{sec1}
It has become common knowledge that data from microarray experiments will contain some non-biological noise. Therefore, the data needs to be adjusted. In this case, implementing the pre-processing will adjust (Huber et al. \cite{Hub04,Hub05a,Hub05b}) or correct the background intensity value. 

There are several steps in pre-processing where one of the steps \color{black} is the background correction. In the background correction, the noise can be modelled as additive or multiplicative (\emph{See}, Huber et al. \cite{Hub04,Hub05a}, Bolstad et al. \cite{Bol03} and Irizarry et al. \cite{Iri03a, Iri03b, Iri06}, Li and Wong \cite{LiW01}, Silver et al. \cite{Sil09} and Wu et al. \cite{Wu04}).

In the robust multi-array average (RMA), Irizarry et al. \cite{Iri03a, Iri03b, Iri06} have modeled the noise as an additive, to adjust the intensity value. Although the RMA was developed for the Affymetrix platform initially, it was also been used for the data from the Illumina platform. 

Currently, there are some models to correct the intensity value of the Illumina platform available, for instance : the model-based background correction method (MBCB) from Ding et al. \cite{Din08} and Xie et al. \cite{Xie09}, the exponential-gamma from Chen et al. \cite{Che11}, the gamma-normal from Plancade et al. \cite{Pla12} and the exponential(gamma)-lognormal from Fajriyah \cite{Faj14}.

Posekany's et al. study \cite{Pos11} show us that by using the Affymetrix and Invitrogen platforms the noise in microarray data is not Gaussian but far more heavy-tailed. On the other hand, Chen et al. \cite{Che11} show that the noise distribution in the Illumina platform is usually skewed in different degrees.

Therefore, while the intensity values are widely accepted as a skewed distribution, the noise distribution could possibly be symmetrical or skewed. Note that in this paper, noise and intensity mean the negative control probes and the observed probes intensity values respectively.

\begin{figure}[!h]
\begin{center}
\includegraphics[width=5in,height=3.25in]{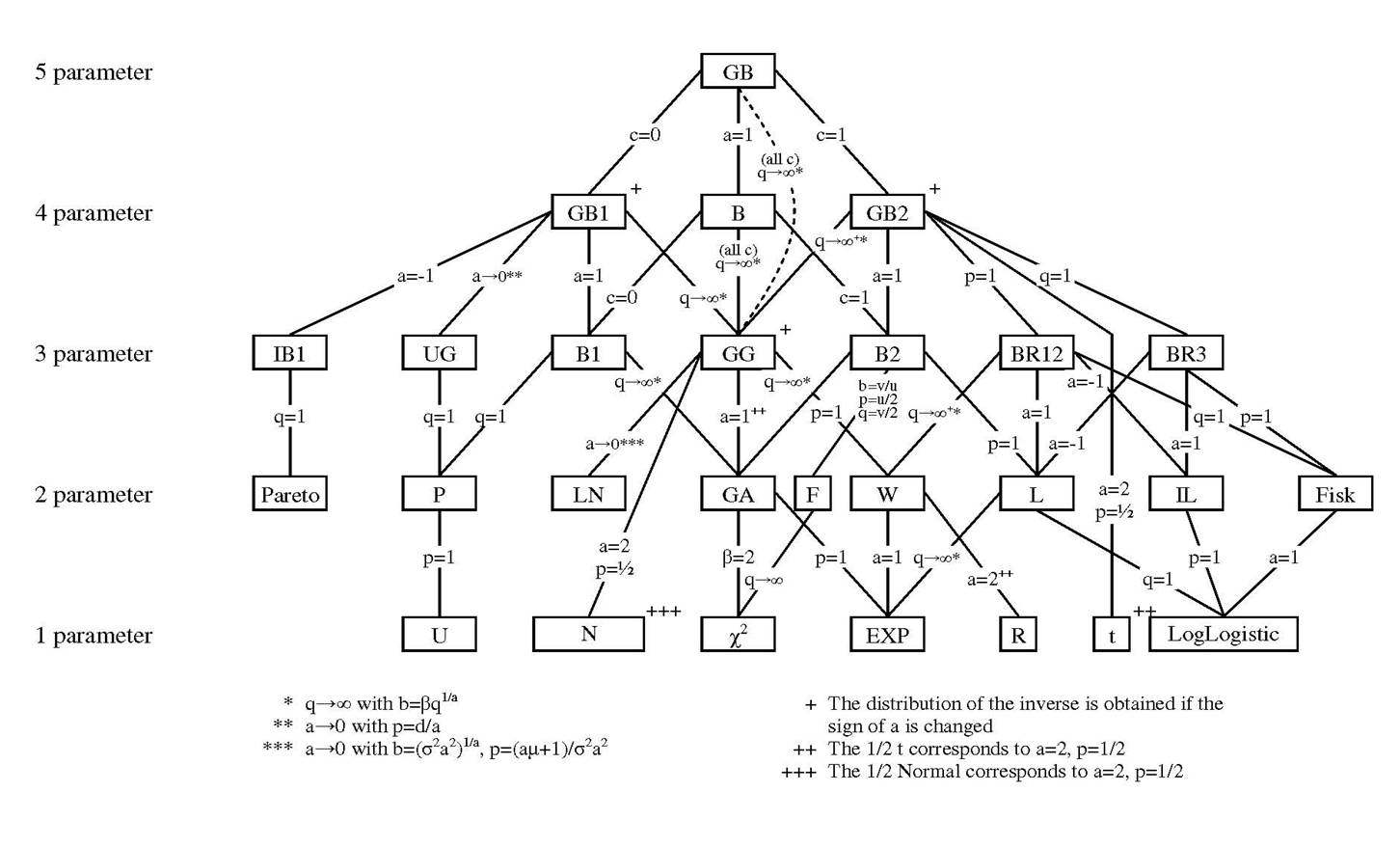}
\end{center}
\captionsetup{font={footnotesize}}
\caption{Distribution tree, \cite{Don95}} \label{fig1}
\end{figure}

McDonald and Xu \cite{Don95} have introduced a distribution tree of generalized beta distributions, which is used to model the income distribution. It is similar in nature to the microarray data where the random variable is a non-negative value. This distribution tree helps us to understand the relationship among the available distributions. Moreover, quite recently, Leemis and McQueston \cite{Lee08} have explained the relationships among the univariate distributions in statistics. See the distribution tree from McDonald and Xu \cite{Don95} in Figure \ref{fig1}.

This paper aims to present the true intensity value, the corrected background intensity, where the noise is a symmetric and skewed distribution. If the noise is a skewed distribution, the underlying distributions of the proposed convolution model are the generalized beta distributions, a generalized model of the existing ones. If the noise is a symmetrically distributed, the proposed model is a generalized beta-normal convolution, which is a generalized model of the Plancade et al. model \cite{Pla12}.

In general, the background correction is applied toward each array, where in each array there are probes (perfect match and mismatch probes), probesets and genes (terminology for the Affymetrix platform) or bead and bead-type level probes (terminology for the Illumina platform).

The current publicly available benchmarking data set for the Illumina platform is the raw data from the bead studio, which is the average of the bead-type level probes, not corrected background and of unnormalized intensity. Therefore, the background correction in this paper is applied to the gene (bead-type level probes) intensity in each array. 

Suppose we have $J$ arrays and for each array there are $I$ regular genes and $W$ negative control genes. Throughout the paper, the convolution model is applied for each array $j$ and represented as follows:
\begin{align} \label{eq11}
P_{i}=S_{i}+B_{i}
\end{align}
where $P_{i}, S_{i}$, and $B_{i}$ are the regular (observed) true/corrected background and noise intensity values respectively of the $i^{\textrm{th}}$ gene, $i=1,...,I$. For a negative control gene $w$ at array $j$, $w=1,2,...,W$, the observed intensity, denoted by $P_{0w}$ is assumed to be $P_{0w}=B_{0w}$, where $B_{0w}$ is the noise intensity. $P_{i}$ and $P_{0w}$ are assumed to be independent.

This paper is organized as follows: Section~\ref{sec2} reviews previous work related to the background correction for the Illumina BeadArrays, Section~\ref{sec3} explains the results of our investigation and Section~\ref{sec4} provides discussion and remarks.
\vspace{1.5cc}
\section{Previous work}\label{sec2}
\subsection{Basic concepts} \label{sec21}
\begin{definition}
Suppose $X$ is a random variable of generalized beta distribution. McDonald and Xu \cite{Don95} define the probability function of the generalized beta distribution as follows
\begin{align} \label{eq21}
GB_{X}(x ;  a,c,d,u,v)= \frac{\abs{a} x^{au -1}\left (1-(1-c)\left(\frac{x}{d}\right)^{a}\right)^{v-1}} {d^{au}B(u,v) \left (1+c\left(\frac{x}{d}\right)^{a} \right)^{u+v}}, 0<x^{a} < \frac{d^{a}}{1-c},
\end{align}
 and zero otherwise, with $B\left( u,v \right)$ is the beta function, $0\le c \le 1$, $a, d, u$ and $v$ positive. 
\end{definition}
\vspace{1.5cc}

\vspace{1.5cc}
\begin{definition}
Let $X$ and $Y$ be two continuous random variables with density functions $f_{1} (x)$ and $f_{2} (y)$ respectively. Assume that both $f_{1} (x)$ and $f_{2} (y)$ are defined for all real numbers. Then the \emph{convolution} $f_{1}*f_{2}$ of $f_{1}$ and $f_{2}$ is the function given by 
\begin{align}\label{eq22}
(f_{1}*f_{2})(z)&=\int \limits_{-\infty}^{+\infty} f_{1} (z-y) f_{2}(y) dy \nonumber \\
&= \int \limits_{-\infty}^{+\infty} f_{2} (z-x) f_{1} (x) dx
\end{align}
\end{definition}
\vspace{1.5cc}
\begin{theorem}
Let $X$ and $Y$ be two independent random variables with density functions  $f_{X}(x)$ and $f_{Y}(y)$ respectively defined for all $x$ and $y$. Then the sum $Z=X+Y$ is a random variable with a density function of $f_{Z}(z)$, where $f_{Z}$ is the convolution of $f_{X}$ and $f_{Y}$.
\end{theorem}

\subsection{Background correction by RMA} \label{sec22}
In the RMA model (\cite{Bol03} and \cite{Iri03a,Iri03b,Iri06}), it is assumed that the intensity values are affected by the noise of the chip. The RMA model is as in the Equation (\ref{eq11}), where $P_{i}=PM_{i}$ is the observed probe level intensity of perfect match probes of the $i^{\textrm{th}}$ gene, $S_{i}$ is the true intensity of the $i^{\textrm{th}}$ gene, with $S_{i} \sim f_{1} (s_{i} ;  \theta_{j}) = \text{Exp}(\theta_{j}), \theta_{j}, s_{i} >0$, and $B_{i}$ is the background noise of the $i^{\textrm{th}}$ gene  with $B_{i} \sim f_{2} (b_{i} ; \mu_{j},\sigma_{j}^{2}) = \mathcal{N}\left(\mu_{j},\sigma_{j}^{2}\right), \mu_{j} \in \mathbb{R}, \sigma_{j}^{2}, b_{i} >0$.

Assuming independence, the joint density of the two-dimensional random variables $\left (S_{i},B_{i} \right)$ is
\begin{align*} %\label{eq23}
f_{S_{i},B_{i}} (s_{i}, b_{i};\mu_{j},\sigma_{j}^{2},\theta_{j})&=\theta_{j} e^{-s_{i} \theta_{j}} f_{2}\left(b_{i};\mu_{j},\sigma_{j}^{2}\right),s_{i}, b_{i} >0\, .
\end{align*}
Furthermore, the transformation formula for two-dimensional densities gives the joint density of $S_{i}$ and $P_{i}$ is
\begin{align} \label{eq24} 
&f_{S_{i},P_{i}} \left (s_{i},p_{i} ; \mu_{j},\sigma_{j}^{2},\theta_{j} \right)\nonumber \\
&=\theta_{j} e^{\left(\frac{\theta_{j}^{2}\sigma_{j}^{2}}{2}-
\left(p_{i}-\mu_{j}\right)\theta_{j}\right)}f_{2}\left(s_{i};p_{i}-\mu_{j}-\sigma_{j}^{2}\theta_{j},\sigma_{j}^{2}\right), 0< s_{i}< p_{i} 
\end{align}

From equation (\ref{eq24}) we get the marginal density of $P_{i}$ and the conditional density of $S_{i}$ given $P_{i}$ in equations (\ref{eq25}) and (\ref{eq26}) below, respectively:
\begin{align} \label{eq25}
f_{P_{i}} \left (p_{i}\right)&=\theta_{j} e^{\left(\frac{\theta_{j}^{2}\sigma_{j}^{2}}{2}-
\left(p_{i}-\mu_{j}\right)\theta_{j}\right)} \left(\Phi \left( \frac{\mu_{S.P,j}}{\sigma_{j}} \right)+\Phi \left( \frac{p_{i}-\mu_{S.P,j}}{\sigma_{j}} \right)  -1\right)
\end{align}
\begin{align} \label{eq26}
f_{S_{i}\mid P_{i}}\left(s_{i} \mid p_{i}\right) &= \frac{f_2 (s_{i}; \mu_{S.P,j}, \sigma_{j}^2)}{\left(\Phi \left( \frac{\mu_{S.P,j}}{\sigma_{j}} \right)+\Phi \left( \frac{p_{i}-\mu_{S.P,j}}{\sigma_{j}} \right)  -1\right)}
\end{align}
where $\mu_{S.P,j}=p_{i}-\mu_{j}-\sigma_{j}^{2} \theta_{j}$.

The corrected background intensity is computed by the conditional expectation
\begin{align} \label{eq27}
E(S_{i}\mid P_{i}=p_{i})&=\frac{1}{\left(\Phi \left( \frac{\mu_{S.P,j}}{\sigma_{j}} \right)+\Phi \left( \frac{p_{i}-\mu_{S.P,j}}{\sigma_{j}} \right)  -1\right)} \int \limits_{0}^{p_{i}} f_{2}\left(s_{i}; \mu_{S.P,j},\sigma_{j}^{2}\right)ds_{i} 
\end{align}

The substitution $s_{i}=\mu_{S.P,j}+\sigma_{j} t_{i}$, yields the corrected background intensity in the Equation (\ref{eq27}) equal to %\cite{Bol03}
\begin{align} \label{eq28}
 &\mu_{S.P,j}+\sigma_{j} \frac{\phi \left (\frac{\mu_{S.P,j}}{\sigma_{j}} \right) - \phi \left (\frac{p_{i}-\mu_{S.P,j}}{\sigma_{j}}\right )}{\Phi \left (\frac{\mu_{S.P,j}}{\sigma_{j}}\right) +\Phi \left (\frac{p_{i}-\mu_{S.P,j}}{\sigma_{j}}\right )-1}
\end{align}
\subsection{Exponential-normal MBCB} \label{sec23}
Xie et al. \cite{Xie09} use the same underlying distributions as the RMA for the background correction. The differences between the MBCB and the RMA (\cite{Bol03} and \cite{Iri03a,Iri03b,Iri06}) are
\begin{enumerate}
\item Xie et al. \cite{Xie09} take the infinite value for the upper bound of the integral to compute the marginal density function and the conditional expectation of the true intensity value. On the other hand, the RMA puts $p$ as the upper bound of the integral. 

The corrected background intensity of this model is %\cite{Xie09}
\begin{eqnarray} \label{eq29}
& \mu_{S.P,j}+\sigma_{j} \frac{\phi \left (\frac{\mu_{S.P,j}}{\sigma_{j}} \right )}{\Phi \left (\frac{\mu_{S.P,j}}{\sigma_{j}} \right )}
\end{eqnarray}
\item Under the convolution model (\ref{eq11}), where the true intensity value is assumed exponentially distributed and the noise is normally distributed, we then need to estimate the parameters $\theta_{j}, \mu_{j},$ and $\sigma_{j}^{2}$. Xie et al. \cite{Xie09}  offer three parameters estimation methods: the non-parametric, maximum likelihood and Bayesian. On the other hand, the RMA applies the \emph{ad-hoc} method.
\end{enumerate} 

Ding et al. \cite{Din08} use the exponential-normal convolution model to correct the background of the Illumina platform by using a Markov chain Monte Carlo simulation.

\subsection{Gamma-normal convolution} \label{sec24}
Plancade et al. \cite{Pla12} introduced gamma-normal convolution to model the background correction of the Illumina BeadArrays. The model is based on the RMA background correction of Affymetrix GeneChips. Plancade et al. \cite{Pla12} assume that the true intensity value is gamma distributed and the noise is normally distributed.

Under the model background correction in (\ref{eq11}), $f_{P_{i}}$ is the convolution product of $f_{S_{i}}$ and $f_{B_{i}}$. The true intensity $S_{i}$ is computed by  the conditional expectation of $S_{i}$ given $P_{i}=p_{i}$:
\begin{align} \label{eq210}
E \left (S_{i}\mid P_{i}=p_{i} \right )&=\tilde{S_{i}}\left (p_{i}\right)= \frac{\int s_{i} f^{\textrm{gam}}_{\alpha_{j},\theta_{j}}(s_{i}) f^{\textrm{norm}}_{\mu_{j},\sigma_{j}}\left (p_{i}-s_{i} \right) ds_{i} }{\int f^{\textrm{gam}}_{\alpha_{j},\theta_{j}}(s_{i}) f^{\textrm{norm}}_{\mu_{j},\sigma_{j}}(p_{i}-s_{i}) ds_{i}}
\end{align}
where $
f_{\alpha_{j}, \theta_{j}}^{\textrm{gam}} \left(x_{i};\alpha_{j}, \theta_{j} \right )=\frac{\theta_{j}^{\alpha_{j}} x_{i}^{\alpha_{j}-1} e^{ -\theta_{j} x_{i}}} {\Gamma \left (\alpha_{j} \right)}, \quad \alpha_{j}, \theta_{j}, x_{i}>0
$  is the gamma density.

When $S_{i}$ is gamma distributed and $B_{i}$ is normally distributed, then the equation (\ref{eq210}) does not have analytic expression as it does in Equations (\ref{eq28}) and (\ref{eq29}). Therefore, Plancade et al. \cite{Pla12} implemented the Fast Fourier Transform to estimate the parameters and to correct the background. For the background correction with Fast Fourier Transform, Equation (\ref{eq210}) is rewritten as
\begin{align} \label{eq211}
\widetilde{S}_{i}(p_{i}\mid \Theta)= \frac{\alpha_{j} \theta_{j} \int f^{\textrm{gam}}_{\alpha_{j}+1,\theta_{j}}\left(s_{i}\right) f^{\textrm{norm}}_{\mu_{j},\sigma_{j}}\left (p_{i}-s_{i} \right) d s_{i} }{\int f^{\textrm{gam}}_{\alpha_{j},\theta_{j}}\left (s_{i} \right ) f^{\textrm{norm}}_{\mu_{j},\sigma_{j}}\left (p_{i}-s_{i} \right ) ds},
\end{align}
where $\Theta=\left (\mu_{j},\sigma_{j},\alpha_{j},\theta_{j} \right )$, %$S_{ij}$ and $B_{ij}$ are independent 
and $s_{i} f_{\alpha_{j},\theta_{j}}^{\textrm{gam}}(s_{i})=\alpha_{j} \theta_{j} f_{\alpha_{j}+1,\theta_{j}}^{\textrm{gam}}\left (s_{i} \right )$ is valid for every $s_{i} > 0$.

\subsection{Exponential-gamma convolution} \label{sec25}
Chen et al. \cite{Che11} proposed in favor of the distribution of the true intensity and its noise, under the convolution model of Equation (\ref{eq11}), the exponential and gamma distributions respectively. Therefore, $S_{i} \sim f_{1} (s_{i} ;  \theta_{j}) = \textrm{Exp} \left (\theta_{j} \right )$, and $B_{i} \sim f_{2} \left (b_{i} ; \alpha_{j}, \beta_{j} \right ) = \textrm{GAM}\left (\alpha_{j}, \beta_{j} \right )$, where $s_{i}, b_{i}, \theta_{j}, \alpha_{j}, \beta_{j} >0$.

The corrected background intensity for the proposed model (\cite{Che11}) is :
\begin{align} \label{eq212}
& p_{i} -\frac{\int \limits_{0}^{p_{i}}  b_{i}^{\alpha_{j}} e^{-\left (\frac{1}{\beta_{j}} -\theta_{j} \right )b_{i}} db_{i}}{ \int \limits_{0}^{p_{i}}  b_{i}^{\alpha_{j}-1} e^{-\left (\frac{1}{\beta_{j}} -\theta_{j} \right ) b_{i}} db_{i}}.
\end{align}
\subsection{Exponential-lognormal convolution, \cite{Faj14}}\label{sec26}
Under  model (\ref{eq11}), when the true intensity $S_{i}$ is assumed to be exponentially distributed $S_{i} \sim f_{1} \left (s_{i} ;  \theta_{j} \right ) = \theta_{j} e^{-\theta_{j} s_{i}}, \theta_{j}, s_{i}  >0$, and the background noise $B$ is assumed to be lognormally distributed, $B_{i} \sim f_{2} (b_{i} ; \mu_{j},\sigma_{j}^{2}) = \frac{e^{-\frac{\left (\ln b_{i} -\mu_{j} \right )^{2}}{2\sigma_{j}^{2}}}}{b_{i} \sigma_{j} \sqrt{2\pi}}, \mu_{j} \in \mathbb{R}, \sigma_{j}^{2}, b_{i} >0$, the corrected background intensity is 
\begin{align} \label{eq213}
&p_{i}-\frac{e^{\mu_{j}+\frac{\sigma_{j}^{2}}{2}}C_{2,j}}{C_{1,j}}  
\end{align}
where \\
$C_{2,j}= \sum \limits_{k=0}^{\infty} \frac{\theta_{j}^{k}}{k!}e^{k \left (\mu_{j}+\frac{k+2}{2}\sigma_{j}^{2} \right )}\Phi \left (\frac{\ln p_{i} - \left (\mu_{j}+(k+1)\sigma_{j}^{2} \right )}{\sigma_{j}} \right )$, and \\ 
$C_{1,j}=\sum \limits_{k=0}^{\infty} \frac{\theta_{j}^{k}}{k!}e^{k \left (\mu_{j}+\frac{k}{2}\sigma_{j}^{2}\right )}\Phi \left (\frac{\ln p_{i} - \left (\mu_{j}+k\sigma_{j}^{2} \right)}{\sigma_{j}}\right)$

\subsection{Gamma-lognormal convolution, \cite{Faj14}}\label{sec27}
Under  model (\ref{eq11}), when the true intensity $S_{i}$ is assumed to be gamma distributed $S_{i} \sim f_{1} (s_{i} ;  \alpha_{j}, \beta_{j}) = \frac{ s_{i}^{\alpha_{j}-1}e^{-\frac{s_{i}}{\beta_{j}}}}{\beta_{j}^{\alpha_{j}} \Gamma \left(\alpha_{j} \right )}, \alpha_{j}, \beta_{j}, s_{i}  >0$, and the background noise $B_{i}$ is assumed to be lognormally distributed, $B_{i} \sim f_{2} (b_{i} ; \mu_{j},\sigma_{j}^{2}) = \frac{e^{-\frac{\left (\ln b_{i} -\mu_{j} \right )^{2}}{2\sigma_{j}^{2}}}}{b_{i} \sigma_{j} \sqrt{2\pi}}, \mu_{j} \in \mathbb{R}, \sigma_{j}^{2}, b_{i} >0$, the corrected background intensity is
\begin{align} \label{eq214}
&\frac{p_{i}C_{4,j}}{C_{3,j}} 
\end{align}
where \\
$C_{4,j}=\sum \limits_{k=0}^{\infty}  \sum_{n=0}^{\infty} \frac{(-1)^{k} \binom{\alpha_{j}}{k} e^{(k+n)\left (\mu_{j} +(k+n)\frac{\sigma_{j}^{2}}{2}\right)}\Phi \left ( \frac{\ln p_{i} -\left (\mu_{j}+(k+n)\sigma_{j}^{2}\right )}{\sigma_{j}}\right)}{p_{i}^{k}\beta_{j}^{n} n!} $, and \\
$C_{3,j}= \sum \limits_{k=0}^{\infty} \sum_{n=0}^{\infty} \frac{(-1)^{k} \binom{\alpha_{j}-1}{k} e^{(k+n)\left (\mu +(k+n)\frac{\sigma_{j}^{2}}{2} \right)}\Phi \left ( \frac{\ln p_{i} -(\mu_{j}+(k+n)\sigma_{j}^{2})}{\sigma_{j}}\right)}{p_{i}^{k}\beta_{j}^{n} n!}$\\

In the exponential-lognormal and gamma-lognormal models, Fajriyah \cite{Faj14}  implements three methods for the parameters estimation: Maximum likelihood estimation (MLE), method of moments, and plug-in.
\vspace{1.5cc}
\section{Results} \label{sec3}
In the subsequent  sections, we will explain the generalized beta convolution model and its corrected background intensity value.

\subsection{Generalized beta distribution convolution}\label{sec31}
\subsubsection{The joint density function}\label{sec311}
Under the convolution model of Equation (\ref{eq11}),  where
$P_{i}$ is the observed intensity of regular probes of the $i^{\textrm{th}}$ gene,
$S_{i}$ is the true intensity of the $i^{\textrm{th}}$ gene, with
\begin{align*} %\label{eq31}
S_{i} &\sim f_{1} \left (s_{i} ;  a_{1,j}, c_{1,j}, d_{1,j}, u_{1,j}, v_{1,j} \right) \nonumber \\
&= \frac{\abs{a_{1,j} } s_{i}^{a_{1,j}u_{1,j} -1}\left (1-\left (1-c_{1,j}\right )\left (\frac{s_{i}}{d_{1,j}}\right )^{a_{1,j}}\right )^{v_{1,j}-1}} {d^{a_{1,j}u_{1,j}}_{1,j} B\left (u_{1,j},v_{1,j} \right ) \left (1+c_{1,j} \left (\frac{s_{i}}{d_{1,j}} \right )^{a_{1,j}} \right )^{u_{1,j}+v_{1,j}}},  \\
&  0\le c_{1,j} \le 1, a_{1,j}, d_{1,j}, u_{1,j} \textrm{ and  } v_{1,j} \textrm{ positive,  } s_{i}>0 \nonumber 
\end{align*}
and $B_{i}$ is the background noise with
\begin{align*} %\label{eq32}
B_{i} &\sim f_{2} \left (b_{i} ; a_{2,j}, c_{2,j}, d_{2,j}, u_{2,j}, v_{2,j} \right ) \nonumber \\
&\frac{\abs{a_{2,j}} b_{i}^{a_{2,j}u_{2,j} -1} \left (1-\left (1-c_{2,j}\right ) \left (\frac{b_{i}}{d_{2,j}} \right )^{a_{2,j}} \right )^{v_{2,j}-1}} {d^{a_{2,j}u_{2,j}}_{2} B\left (u_{2,j},v_{2,j} \right ) \left (1+c_{2,j} \left (\frac{b_{i}}{d_{2,j}} \right )^{a_{2,j}} \right )^{u_{2,j}+v_{2,j}}}, \\
& 0 \le c_{2,j} \le 1,  a_{2,j}, d_{2,j}, u_{2,j} \textrm{ and  } v_{2,j} \textrm{ positive,  } b_{i}  > 0 \nonumber 
\end{align*} 
The joint density function of $S_{i}$ and $B_{i}$ is :
\begin{align*} %\label{eq33}
f_{S_{i},B_{i}} \left (s_{i},b_{i} \right )&=\frac{\abs{a_{1}} s_{i}^{a_{1,j}u_{1,j} -1}\left (1- \left (1-c_{1,j} \right ) \left (\frac{s_{i}}{d_{1}} \right )^{a_{1,j}}\right )^{v_{1,j}-1}} {d^{a_{1,j}u_{1,j}}_{1} B\left (u_{1,j},v_{1,j} \right ) \left(1+c_{1,j} \left (\frac{s_{i}}{d_{1,j}} \right )^{a_{1,j}} \right )^{u_{1,j}+v_{1,j}}} \times \nonumber \\
& \frac{\abs{a_{2,j}} b_{i}^{a_{2,j}u_{2,j} -1} \left (1- \left (1-c_{2,j} \right ) \left (\frac{b_{i}}{d_{2,j}}\right )^{a_{2,j}} \right )^{v_{2,j}-1}} {d^{a_{2,j}u_{2,j}}_{2} B \left (u_{2,j},v_{2,j} \right ) \left (1+c_{2,j} \left (\frac{b_{i}}{d_{2,j}} \right )^{a_{2,j}} \right )^{u_{2,j}+v_{2,j}}}
\end{align*}
The joint density function of $S_{i}$ and $P_{i}$ is
\begin{align*} %\label{eq34}
f_{S_{i},P_{i}} \left (s_{i},p_{i} \right)&= \frac{\abs{a_{1}} s_{i}^{a_{1,j}u_{1,j} -1} \left (1-\left(1-c_{1,j} \right ) \left (\frac{s_{i}}{d_{1,j}}\right)^{a_{1,j}} \right)^{v_{1,j}-1}} {d^{a_{1,j}u_{1,j}}_{1} B \left (u_{1},v_{1,j} \right ) \left(1+c_{1,j} \left (\frac{s_{i}}{d_{1,j}}\right )^{a_{1,j}}\right )^{u_{1,j}+v_{1,j}}} \times \nonumber \\
& \frac{\abs{a_{2,j}} \left (p_{i}-s_{i} \right )^{a_{2,j}u_{2,j} -1}\left (1-\left (1-c_{2,j} \right ) \left (\frac{ \left (p_{i}-s_{i} \right )}{d_{2,j}} \right )^{a_{2,j}}\right )^{v_{2,j}-1}} {d^{a_{2,j}u_{2,j}}_{2} B \left (u_{2,j},v_{2,j} \right ) \left (1+c_{2,j}\left (\frac{\left (p_{i}-s_{i} \right )}{d_{2,j}} \right )^{a_{2,j}} \right)^{u_{2,j}+v_{2,j}}}
\end{align*}

\subsubsection{The marginal density function} \label{sec312}
The marginal density function of $P_{i}$ is
\begin{small}
\begin{align} \label{eq35}
f_{P_{i}} \left (p_{i}\right)&= \int \limits_{0}^{p_{i}} f_{S_{i},P_{i}}\left (s_{i},p_{i} \right) ds_{i} \nonumber \\
&= K \sum \limits_{l=0}^{\infty} \sum_{m=0}^{\infty} \sum_{n=0}^{\infty}\sum_{r=0}^{\infty} \Bigg \{\frac{(-1)^{l+m+n+r} \left (1-c_{1,j} \right)^{l} \left (1-c_{2,j} \right )^{m} c_{1,j}^{n} c_{2,j}^{r} } {d_{1,j}^{a_{1,j}(l+n)}d_{2,j}^{a_{2,j}(m+r)}} \times \nonumber \\
& \dbinom{v_{1,j}-1}{l} \dbinom{v_{2,j}-1}{m} \dbinom{u_{1,j}+v_{1,j}+n-1}{n} \dbinom{u_{2,j}+v_{2,j}+r-1}{r} \times \nonumber \\
&\int \limits_{0}^{p_{i}} s_{i}^{a_{1,j}(u_{1,j} +l+n)-1} \left (p_{i}-s_{i} \right )^{a_{2,j} \left (u_{2,j}+m+r \right ) -1}ds_{i} \Bigg \} 
\end{align}
\end{small}
Let $\frac{s_{i}}{p_{i}} = z_{i}$, then the equation (\ref{eq35}) becomes
\begin{align} \label{eq36}
 K_{1} p_{i}^{a_{1,j}u_{1,j} +a_{2,j}u_{2,j}-1} C_{5,j}
\end{align}
where
\begin{small}
\begin{align} \nonumber
K_{1}&=\frac{\mid a_{1,j} \mid \mid a_{2,j} \mid } {d^{a_{1,j}u_{1,j}}_{1} d^{a_{2,j}u_{2,j}}_{2} B(u_{1,j},v_{1,j}) B(u_{2,j},v_{2,j})},
\nonumber
\end{align}  and
\begin{align} \nonumber
C_{5,j}&=   \sum \limits_{l=0}^{\infty} \sum_{m=0}^{\infty} \sum_{n=0}^{\infty}\sum_{r=0}^{\infty}  \Bigg \{\frac{(-1)^{l+m+n+r} \left (1-c_{1,j} \right )^{l} \left (1-c_{2,j} \right )^{m}c_{1,j}^{n} c_{2,j}^{r} \dbinom{v_{1,j}-1}{l} \dbinom{v_{2,j}-1}{m}} {d_{1,j}^{a_{1,j}(l+n)}d_{2,j}^{a_{2,j}(m+r)}} \times \nonumber \\
&   \dbinom{u_{1,j}+v_{1,j}+n-1}{n} \dbinom{u_{2,j}+v_{2,j}+r-1}{r} p_{i}^{a_{1,j}(l+n)+a_{2,j}(m+r)} \times \nonumber \\
& \textrm{B} \left (a_{1,j} \left (u_{1,j} +l+n\right )-1,a_{2,j}\left (u_{2,j}+m+r \right) -1\right) \Bigg \} \nonumber 
\end{align}
\end{small}

\subsubsection{The conditional density function} \label{sec313}
The conditional density function of $S_{i}$ where it is known that $P_{i}=p_{i}$ is
\begin{align*} %\label{eq37}
f_{S_{i}\mid P_{i}}\left(s_{i} \mid p_{i}\right) &=\frac{f_{S_{i},P_{i}} \left(s_{i},p_{i}\right )}{f_{P_{i}}\left (p_{i} \right)} \nonumber \\
&=\frac{ s_{i}^{a_{1,j}u_{1,j} -1}\left (1- \left (1-c_{1,j} \right ) \left (\frac{s_{i}}{d_{1,j}}\right )^{a_{1,j}} \right )^{v_{1,j}-1} \left (p_{i}-s_{i} \right )^{a_{2,j}u_{2,j} -1}}{p_{i}^{a_{1,j}u_{1,j} +a_{2,j}u_{2,j}-1} C_{5,j}\left (1+c_{1,j} \left (\frac{s_{i}}{d_{1,j}} \right)^{a_{1,j}} \right )^{u_{1,j}+v_{1,j}}} \times \nonumber \\
&
\frac{\left (1-\left(1-c_{2,j}\right) \left(\frac{\left(p_{i}-s_{i} \right)}{d_{2,j}}\right)^{a_{2,j}}\right)^{v_{2,j}-1}} { \left(1+c_{2,j}\left(\frac{\left(p_{i}-s_{i}\right)}{d_{2,j}}\right)^{a_{2,j}} \right)^{u_{2,j}+v_{2,j}} }
\end{align*}

\subsubsection{The corrected background intensity} \label{sec314}
The corrected background intensity under this generalized beta convolution is 
\begin{align} \label{eq38}
& p_{i}\frac{C_{6,j}}{C_{5,j}}
\end{align}
where  \begin{small}
\begin{align*}
C_{6,j}=& \sum \limits_{l=0}^{\infty} \sum_{m=0}^{\infty} \sum_{n=0}^{\infty}\sum_{r=0}^{\infty}  \left \{ \frac{(-1)^{l+m+n+r} \left (1-c_{1,j} \right)^{l} \left(1-c_{2,j}\right)^{m} c_{1,j}^{n} c_{2,j}^{r}  \dbinom{v_{1,j}-1}{l} \dbinom{v_{2,j}-1}{m}  } {d_{1,j}^{a_{1,j}(l+n)}d_{2,j}^{a_{2,j}(m+r)}} \times \right. \nonumber \\
&  \left.  \dbinom{u_{1,j}+v_{1,j}+n-1}{n}  \dbinom{u_{2,j}+v_{2,j}+r-1}{r} p_{i}^{a_{1,j}(l+n)+a_{2,j}(m+r)}  \times \right. \nonumber \\
&  \left. \textrm{B} \left (a_{1,j} \left(u_{1,j} +l+n\right),a_{2,j}\left(u_{2,j}+m+r\right) -1 \right ) \right \} \nonumber
\end{align*}
\end{small}

\subsubsection{The likelihood function} \label{sec315}
The likelihood function ($\textbf{L}$) to estimate  $ a_{1,j}, c_{1,j}, d_{1,j}, u_{1,j}, v_{1,j},  a_{2,j}, c_{2,j}, d_{2,j}, u_{2,j}$ and $v_{2,j}$ is
\begin{align*} %\label{eq39}
=&\prod \limits_{i=1}^{I}\frac{ \abs{a_{1,j}} \abs{a_{2,j}} p_{i}^{a_{1,j}u_{1} +a_{2,j}u_{2,j}-1} C_{5,j}} {d^{a_{1,j}u_{1,j}}_{2} d^{a_{2,j}u_{2,j}}_{2,j} \textrm{B}\left(u_{1},v_{1,j}\right) B\left(u_{2,j},v_{2,j}\right)} \times \nonumber \\
&\prod \limits_{w=1}^{W}\frac{\abs{a_{2,j}} b_{0w}^{a_{2,j}u_{2,j} -1}\left(1-\left(1-c_{2,j}\right) \left(\frac{b_{0w}}{d_{2,j}}\right)^{a_{2,j}}\right)^{v_{2,j}-1}} {d^{a_{2,j}u_{2,j}}_{2} B\left(u_{2,j},v_{2,j} \right) \left(1+c_{2,j}\left (\frac{b_{0w}}{d_{2,j}} \right)^{a_{2,j}} \right)^{u_{2,j}+v_{2,j}}} 
\end{align*}
The log-likelihood function $l$ is
\begin{align} \label{eq310}
= & \sum \limits_{i=1}^{I} \left \{ \ln \left(\abs{a_{1,j}} \right)+ \ln \left (\mid a_{2,j}\mid \right)+\left(a_{1,j}u_{1,j} + a_{2,j}u_{2,j} -1\right) \ln \left(p_{i}\right)\right. \nonumber \\
&  \left. +\ln \left(C_{5,j}\right) -\left(a_{1,j}u_{1,j}\right) \ln \left(d_{1,j}\right) - \left(a_{2,j}u_{2,j}\right) \ln \left(d_{2,j} \right) -\ln \left(\textrm{B} \left(u_{1,j},v_{1,j}\right)\right) \right. \nonumber \\
&  \left.  -\ln \left(\textrm{B} \left(u_{2,j},v_{2,j}\right) \right) \right \} + \sum \limits_{w=1}^{W} \left \{ \ln \left(\mid a_{2,j} \mid \right) +\left(a_{2,j}u_{2,j}-1\right) \ln \left(b_{0w}\right) \right. \nonumber \\
&  \left.  + \left(v_{2,j}-1\right) \ln \left ( \left(1-\left (1-c_{2,j} \right) \left(\frac{b_{0w}}{d_{2,j}}\right)^{a_{2,j}}\right) \right)- \left(a_{2,j}u_{2,j} \right)\ln \left(d_{2,j}\right) \right. \nonumber \\
&  \left.  - \ln \left(\textrm{B} \left(u_{2,j},v_{2,j}\right)\right) - \left(u_{2,j}+v_{2,j}\right)\ln \left( \left(1+c_{2,j}\left(\frac{b_{0w}}{d_{2,j}}\right)^{a_{2,j}} \right) \right) \right \}
\end{align}
%\color{blue}
The likelihood equations are as follows
\begin{align*} %\label{eq311}
\frac{\partial l}{\partial a_{2,j}} = & \sum \limits_{w=1}^{W} \left ( \frac{1}{\abs{a_{2,j}}} + u_{2,j} \ln \left( b_{0w} \right) + \left( v_{2,j} -1 \right ) \frac{-\left(1-c_{2,j} \right)  \ln \left( \frac{b_{0w}}{d_{2,j}} \right)\left( \frac{b_{0w}}{d_{2,j}}\right)^{a_{2,j}}}{\left(1-\left(1-c_{2,j} \right) \left(\frac{b_{0w}}{d_{2,j}}\right)^{a_{2,j}} \right)} \right. \nonumber \\
&  \left. - u_{2,j} \ln \left( d_{2,j} \right) -  \left( u_{2,j} + v_{2,j} \right) \frac{c_{2,j} \ln \left( \frac{b_{0w}}{d_{2,j}} \right)\left( \frac{b_{0w}}{d_{2,j}}\right)^{a_{2,j}}}{1+c_{2,j} \left( \frac{b_{0w}}{d_{2,j}}\right)^{a_{2,j}}}  \right )=0
\end{align*}
\begin{align*} %\label{eq312}
\frac{\partial l}{\partial c_{2,j}} = & \sum \limits_{w=1}^{W} \left ( \left( v_{2,j} -1 \right ) \frac{ \left(\frac{b_{0w}}{d_{2,j}}\right)^{a_{2,j}}}{\left(1-\left(1-c_{2,j} \right) \left(\frac{b_{0w}}{d_{2,j}}\right)^{a_{2,j}} \right)}  \right. \nonumber \\
&  \left. - \left(u_{2,j} + v_{2,j} \right ) \frac{ \left(\frac{b_{0w}}{d_{2,j}}\right)^{a_{2,j}}}{\left(1+c_{2,j} \left(\frac{b_{0w}}{d_{2,j}}\right)^{a_{2,j}} \right)}  \right ) =0
\end{align*}
\begin{align*} %\label{eq313}
\frac{\partial l}{\partial d_{2,j}} = & \sum \limits_{w=1}^{W} \left ( \left( v_{2,j} -1 \right ) \frac{-\left(1-c_{2,j} \right) b_{0w}^{a_{2,j}} \left( -a_{2,j} \right) d_{2,j}^{-\left( a_{2,j}+1 \right)}}{\left(1-\left(1-c_{2,j} \right) \left(\frac{b_{0w}}{d_{2,j}}\right)^{a_{2,j}} \right)} - \frac{a_{2,j} u_{2,j}}{d_{2,j}}  \right. \nonumber \\
&  \left. - \left( u_{2,j}+v_{2,j} \right) \frac{c_{2,j} b_{0w}^{a_{2,j}} \left(-a_{2,j} \right) d_{2,j}^{-\left(a_{2,j}+1 \right)}}{\left(1+c_{2,j} \left(\frac{b_{0w}}{d_{2,j}}\right)^{a_{2,j}} \right)}  \right )=0
\end{align*}
\begin{align*} %\label{eq314}
\frac{\partial l}{\partial u_{2,j}} = & \sum \limits_{w=1}^{W} \left ( a_{2,j} \ln \left(b_{0w} \right) + a_{2,j} \ln \left (d_{2,j} \right) - \frac{\frac{\partial \textrm{B} \left(u_{2,j}, v_{2,j} \right) }{\partial u_{2,j}}}{\textrm{B} \left(u_{2,j}, v_{2,j} \right) } \right. \nonumber \\
&  \left.  - \ln \left(1+c_{2,j} \left(\frac{b_{0w}}{d_{2,j}}\right)^{a_{2,j}} \right)  \right )=0
\end{align*}
\begin{align*} %\label{eq315}
\frac{\partial l}{\partial v_{2,j}} = & \sum \limits_{w=1}^{W} \left ( \ln \left(1-\left(1-c_{2,j} \right) \left(\frac{b_{0w}}{d_{2,j}}\right)^{a_{2,j}} \right) - \frac{\frac{\partial \textrm{B} \left(u_{2,j}, v_{2,j} \right) }{\partial v_{2,j}}}{\textrm{B} \left(u_{2,j}, v_{2,j} \right) } \right. \nonumber \\
&  \left.  - \ln \left(1+c_{2,j} \left(\frac{b_{0w}}{d_{2,j}}\right)^{a_{2,j}} \right) \right ) =0
\end{align*}
\begin{align*} %\label{eq316}
\frac{\partial l}{\partial a_{1,j}} = & \sum \limits_{i=1}^{I} \left ( \frac{1}{\abs{a_{1,j}}} + u_{1,j} \ln \left( p_{i} \right) + \frac{\frac{\partial C_{5,j}}{\partial a_{1,j}}}{C_{5,j}} - u_{1,j} \ln \left( d_{1,j} \right ) \right)=0
\end{align*}
\begin{align*} %\label{eq317}
\frac{\partial l}{\partial c_{1,j}} = & \sum \limits_{i=1}^{I} \left ( \frac{\frac{\partial C_{5,j}}{\partial c_{1,j}}}{C_{5,j}} \right)=0
\end{align*}
\begin{align*} %\label{eq318}
\frac{\partial l}{\partial d_{1,j}} = & \sum \limits_{i=1}^{I} \left ( \frac{\frac{\partial C_{5,j}}{\partial d_{1,j}}}{C_{5,j}} -\frac{a_{1,j}u_{1,j}}{d_{1,j}} \right) = 0
\end{align*}
\begin{align*} %\label{eq319}
\frac{\partial l}{\partial u_{1,j}} = & \sum \limits_{i=1}^{I} \left ( a_{1,j} \ln \left(p_{i} \right) + \frac{\frac{\partial C_{5,j}}{\partial u_{1,j}}}{C_{5,j}} -a_{1,j} \ln \left(d_{1,j} \right)- \frac{\frac{\partial \textrm{B} \left(u_{1,j}, v_{1,j} \right) }{\partial u_{1,j}}}{\textrm{B} \left(u_{1,j}, v_{1,j} \right) }  \right) = 0
\end{align*}
\begin{align*} %\label{eq320}
\frac{\partial l}{\partial v_{1,j}} = & \sum \limits_{i=1}^{I} \left (\frac{\frac{\partial C_{5,j}}{\partial v_{1,j}}}{C_{5,j}} - \frac{\frac{\partial \textrm{B} \left(u_{1,j}, v_{1,j} \right) }{\partial v_{1,j}}}{\textrm{B} \left(u_{1,j}, v_{1,j} \right) } \right) = 0
\end{align*}
where
\begin{align*} %\label{eq321}
\frac{\partial \textrm{B} \left(u_{2,j}, v_{2,j} \right) }{\partial u_{2,j}} &= \Gamma \left(v_{2,j} \right ) \frac{\Gamma\left(u_{2,j}\right) \left(-\gamma + \sum \limits_{k=1}^{u_{2,j}-1} \frac{1}{k} \right) - \Gamma\ \left(u_{2,j} \right) \left( -\gamma + \sum \limits_{k=1}^{u_{2,j}+v_{2,j}-1} \frac{1}{k}\right)}{\Gamma \left( u_{2,j}+v_{2,j} \right)} \nonumber \\
&= \textrm{B} \left(u_{2,j}, v_{2,j} \right) \left( \sum \limits_{k=1}^{u_{2,j}-1} \frac{1}{k} - \sum \limits_{k=1}^{u_{2,j}+v_{2,j}-1} \frac{1}{k} \right)
\end{align*}
\begin{align*} %\label{eq322}
\frac{\partial \textrm{B} \left(u_{2,j}, v_{2,j} \right) }{\partial v_{2,j}} &= \Gamma \left(u_{2,j} \right ) \frac{\Gamma\left(v_{2,j}\right) \left(-\gamma + \sum \limits_{k=1}^{v_{2,j}-1} \frac{1}{k} \right) - \Gamma\ \left(v_{2,j} \right) \left( -\gamma + \sum \limits_{k=1}^{u_{2,j}+v_{2,j}-1} \frac{1}{k}\right)}{\Gamma \left( u_{2,j}+v_{2,j} \right)} \nonumber \\
&= \textrm{B} \left(u_{2,j}, v_{2,j} \right) \left( \sum \limits_{k=1}^{v_{2,j}-1} \frac{1}{k} - \sum \limits_{k=1}^{u_{2,j}+v_{2,j}-1} \frac{1}{k} \right)
\end{align*}
and suppose $C_{5,j}$ is written as $\sum \limits_{l=0}^{\infty} \sum \limits_{m=0}^{\infty} \sum \limits_{n=0}^{\infty} \sum \limits_{r=0}^{\infty} C_{5lmnr}$ then
\begin{align*} %\label{eq323}
\frac{\partial C_{5,j}}{\partial a_{1,j}}=& \sum \limits_{l=0}^{\infty} \sum \limits_{m=0}^{\infty} \sum \limits_{n=0}^{\infty} \sum \limits_{r=0}^{\infty} \left [ C_{5lmnr} \Bigg ( \left( l+n \right) \ln \left ( \frac{p_{i}}{d_{1,j}}\right) + \left( u_{1,j}+l+n \right) \times \right. \nonumber \\
& \left. \left ( \sum \limits_{k=1}^{a_{1,j} \left( u_{1,j}+l+n\right) -1} \frac{1}{k}- \sum \limits_{k=1}^{a_{1,j} \left( u_{1,j}+l+n\right)-a_{2,j} \left( v_{2,j}+m+r \right) -2} \frac{1}{k} \right) \Bigg ) \right ]
\end{align*}
\begin{align*} %\label{eq324}
&\frac{\partial C_{5,j}}{\partial c_{1,j}} = \sum \limits_{l=0}^{\infty} \sum \limits_{m=0}^{\infty} \sum \limits_{n=0}^{\infty}\sum \limits_{r=0}^{\infty} \left [ C_{5lmnr} \left (\frac{\left(1 - c_{1,j} \right)n-lc_{1,j}}{c_{1,j}\left(1 - c_{1,j} \right)} \right) \right ]
\end{align*}
\begin{align*} %\label{eq325}
&\frac{\partial C_{5,j}}{\partial d_{1,j}}  = \sum \limits_{l=0}^{\infty} \sum \limits_{m=0}^{\infty} \sum \limits_{n=0}^{\infty} \sum \limits_{r=0}^{\infty}\left [ C_{5lmnr} \left (\frac{-a_{1,j} \left(l+n\right)}{d_{1,j}} \right ) \right ]
\end{align*}
\begin{align*} %\label{eq326}
\frac{\partial C_{5,j}}{\partial u_{1,j}} =& \sum \limits_{l=0}^{\infty} \sum \limits_{m=0}^{\infty} \sum \limits_{n=0}^{\infty} \sum \limits_{r=0}^{\infty}\left [ C_{5lmnr} \Bigg ( \left ( \sum \limits_{k=1}^{u_{1,j}+v_{1,j}+n-1} \frac{1}{k}- \sum \limits_{k=1}^{u_{1,j}+v_{1,j}-1} \frac{1}{k} \right) + \right. \nonumber \\
& \left. a_{1,j} \left ( \sum \limits_{k=1}^{a_{1,j} \left( u_{1,j}+l+n\right)-1} \frac{1}{k} - \sum \limits_{k=1}^{a_{1,j} \left( u_{1,j}+l+n\right)+a_{2,j} \left( v_{2,j}+m+r\right)-2} \frac{1}{k}\right) \Bigg ) \right ]
\end{align*}
\begin{align*} %\label{eq327}
\frac{\partial C_{5,j}}{\partial v_{1,j}} =& \sum \limits_{l=0}^{\infty} \sum \limits_{m=0}^{\infty} \sum \limits_{n=0}^{\infty} \sum \limits_{r=0}^{\infty} \left [ C_{5lmnr} \Bigg ( \left ( \sum \limits_{k=1}^{v_{1,j}-1} \frac{1}{k}- \sum \limits_{k=1}^{v_{1,j}-l-1} \frac{1}{k} \right) + \right. \nonumber \\
& \left.  \left ( \sum \limits_{k=1}^{u_{1,j}+v_{1,j}+n-1} \frac{1}{k} - \sum \limits_{k=1}^{ u_{1,j}+v_{1,j}-1} \frac{1}{k}\right) \Bigg ) \right ]
\end{align*}
\begin{align*} %\label{eq328}
\frac{\partial \textrm{B} \left(u_{1,j}, v_{1,j} \right) }{\partial u_{1,j}} &= \Gamma \left(v_{1,j} \right ) \frac{\Gamma\left(u_{1,j}\right) \left(-\gamma + \sum \limits_{k=1}^{u_{1,j}-1} \frac{1}{k} \right) - \Gamma\ \left(u_{1,j} \right) \left( -\gamma + \sum \limits_{k=1}^{u_{1,j}+v_{1,j}-1} \frac{1}{k}\right)}{\Gamma \left( u_{1,j}+v_{1,j} \right)} \nonumber \\
&=\textrm{B} \left(u_{1,j}, v_{1,j} \right) \left(\sum \limits_{k=1}^{u_{1,j}-1} \frac{1}{k} - \sum \limits_{k=1}^{u_{1,j}+v_{1,j}-1} \frac{1}{k}\right)
\end{align*}
\begin{align*} %\label{eq329}
\frac{\partial \textrm{B} \left(u_{1,j}, v_{1,j} \right) }{\partial v_{1,j}} &= \Gamma \left(u_{1,j} \right ) \frac{\Gamma\left(v_{1,j}\right) \left(-\gamma + \sum \limits_{k=1}^{v_{1,j}-1} \frac{1}{k} \right) - \Gamma\ \left(v_{1,j} \right) \left( -\gamma + \sum \limits_{k=1}^{u_{1,j}+v_{1,j}-1} \frac{1}{k}\right)}{\Gamma \left( u_{1,j}+v_{1,j} \right)} \nonumber \\
&=\textrm{B} \left(u_{1,j}, v_{1,j} \right) \left(\sum \limits_{k=1}^{v_{1,j}-1} \frac{1}{k} - \sum \limits_{k=1}^{u_{1,j}+v_{1,j}-1} \frac{1}{k}\right)
\end{align*}
and $\quad \gamma$ is the Euler-Mascheroni constant.
%\color{black}
\subsection{Generalized beta-normal convolution}\label{sec32}
Although Figure \ref{fig1} covers normal distribution, we can not derive the formula of the true intensity value when the noise is normal, from Equation (\ref{eq11}). The normal distribution in Figure \ref{fig1} is the normal distribution with one parameter. Therefore, in this section, we derive the formula to compute the corrected background intensity when the noise is symmetrically distributed, a normal distribution. 

\subsubsection{The joint density function}\label{sec321}
Under the convolution model in Equation (\ref{eq11}),  where
$P_{i}$ is the observed intensity of the regular $i^{\textrm{th}}$ gene,
$S_{i}$ is the true intensity of the $i^{\textrm{th}}$ gene, with 
\begin{align*} %\label{eq330}
S_{i} \sim& f_{1} \left (s_{i} ;  a_{j}, c_{j}, d_{j}, u_{j}, v_{j} \right) \nonumber \\
=& \frac{\abs{a_{j}} s_{i}^{a_{j}u_{j}-1}\left (1- \left(1-c_{j} \right) \left(\frac{s_{i}}{d_{j}}\right)^{a_{j}}\right)^{v_{j}-1}} {d^{a_{j}u_{j}} \textrm{B}\left(u_{j},v_{j}\right) \left(1+c_{j}\left(\frac{s_{i}}{d_{j}}\right)^{a_{j}} \right)^{u_{j}+v_{j}}},  \\
& 0\le c_{j} \le 1;  a_{j}, d_{j}, u_{j}, v_{j}, s_{i} > 0 \nonumber 
\end{align*}
and $B$ is the background noise with 
\begin{align*} %\label{eq331}
B_{i}& \sim f_{2} \left (b_{i} ; \mu_{j}, \sigma_{j}^{2} \right) = \frac{e^{-\frac{1}{2\sigma_{j}^{2}}\left(b_{i}-\mu_{j}\right)^{2}}} {\sqrt{2\pi} \sigma_{j}}, \mu_{j} \in \mathbb{R}, \sigma_{j}^{2} >0,  b_{i} >0
\end{align*} 
The joint density function of $S_{i}$ and $B_{i}$ is
\begin{align*} %\label{eq332}
f_{S_{i},B_{i}} \left(s_{i},b_{i} \right)&=\frac{\abs{a_{j}} s_{i}^{a_{j}u_{j}-1} \left(1-\left(1-c_{j}\right)\left(\frac{s_{i}}{d_{j}}\right)^{a_{j}}\right)^{v_{j}-1}} {d_{j}^{a_{j}u_{j}} \textrm{B}\left(u_{j},v_{j}\right) \left(1+c_{j} \left(\frac{s_{i}}{d_{j}}\right)^{a_{j}}\right)^{u_{j}+v_{j}}}\frac{e^{-\frac{1}{2\sigma_{j}^{2}}\left (b_{i}-\mu_{j} \right)^{2}}} {\sqrt{2\pi} \sigma_{j}}
\end{align*}
The joint density function of $S_{i}$ and $P_{i}$ is
\begin{align*} %\label{eq333}
f_{S_{i},P_{i}} \left(s_{i},p_{i} \right) &=\frac{\abs{ a_{j}} s_{i}^{a_{j}u_{j}-1}\left (1-\left (1-c_{j}\right )\left(\frac{s_{i}}{d_{j}} \right)^{a_{j}}\right)^{v_{j}-1}} {d_{j}^{a_{j}u_{j}} \textrm{B}\left(u_{j},v_{j}\right)\left(1+c_{j}\left (\frac{s_{i}}{d_{j}} \right)^{a_{j}}\right)^{u_{j}+v_{j}}}\frac{e^{-\frac{\left (p_{i}-s_{i}-\mu_{j} \right)^{2}}{2\sigma_{j}^{2}}}} {\sqrt{2\pi} \sigma_{j}}
\end{align*}

\subsubsection{The marginal density function} \label{sec322}
The marginal density function of $P_{i}$ is
\begin{align} \label{eq334}
f_{P_{i}}\left(p_{i}\right)=& \frac{\abs{a_{j}}}{d_{j}^{a_{j}u_{j}}\textrm{B}\left(u,v\right) \sqrt{2\pi}\sigma_{j}}\sum \limits_{l=0}^{\infty} \sum_{m=0}^{\infty} \Bigg \{\frac{(-1)^{l+m} \left(1-c_{j}\right)^{l} c_{j}^{m} \dbinom{v_{j}-1}{l} } {d^{a_{j}(l+m)}} \times \nonumber \\
& \dbinom{u_{j}+v_{j}+m-1}{m} \int \limits_{0}^{p_{i}} s_{i}^{a_{j}(u_{j} +l+m)-1} e^{-\frac{(s_{i}-p_{i}-\mu_{j})^{2}}{2\sigma_{j}^{2}}} ds_{i} \Bigg \}
\end{align}
Let $\frac{\left(s_{i}-\left(p_{i}-\mu_{j}\right) \right)}{\sigma_{j}} = z_{i}$, and the equation (\ref{eq334}) becomes 
\begin{align} \label{eq335}
=& \frac{\abs{a_{j}}}{d_{j}^{a_{j}u_{j}}\textrm{B}\left(u_{j},v_{j}\right) \sqrt{2\pi}} \sum \limits_{l=0}^{\infty} \sum_{m=0}^{\infty} \sum_{n=0}^{\infty} \Bigg \{\frac{(-1)^{l+m} \left (1-c_{j} \right)^{l} c_{j}^{m} \dbinom{v_{j}-1}{l} } {d_{j}^{a_{j}(l+m)} \left(p_{i}-\mu_{j}\right)^{n}} \times \nonumber \\
&  \dbinom{u_{j}+v_{j}+m-1}{m} \dbinom{a_{j}\left (u_{j}+l+m\right)-1}{n} \left(p_{i}-\mu_{j}\right)^{a_{j}\left(u_{j}+l+m\right)-1}  \sigma_{j}^{n}\times \nonumber \\
& \int \limits_{-\frac{\left(p_{i}-\mu_{j}\right)}{\sigma_{j}}}^{\frac{\mu_{j}}{\sigma_{j}}} z_{i}^{n} e^{-\frac{z_{i}^{2}}{2}} dz_{i} \Bigg \} 
\end{align}
Let $\frac{z_{i}^{2}}{2} = x_{i}$, the equation (\ref{eq335}) becomes
\begin{align} \label{eq336}
K_{2} C_{7,j}
\end{align}
where
\begin{align} \nonumber 
K_{2}=&\frac{\abs{a_{j}} p_{i}^{a_{j}u_{j}-1}} {2\sqrt{\pi} d_{j}^{a_{j}u_{j}} B\left(u_{j},v_{j}\right) },  \nonumber \\
C_{7,j} = &\displaystyle \sum_{l=0}^{\infty} \sum_{m=0}^{\infty} \sum_{n=0}^{\infty} \Bigg \{ \frac{ (-1)^{l+m} \left(1-c_{j}\right)^{l} c_{j}^{m} \dbinom{v_{j}-1}{l}\dbinom{u_{j}+v_{j}+m-1}{m}  } {d_{j}^{a_{j}(l+m)} \left(p_{i}-\mu_{j} \right)^{n}} \times \nonumber \\
& \dbinom{a_{j}\left (u_{j}+l+m \right)-1}{n} \left(p_{i}-\mu_{j}\right)^{a_{j}(l+m)}  \sigma_{j}^{n} 2^{\frac{n}{2}} 
\Bigg (\gamma \left(\frac{n+1}{2}, \left (\frac{\mu_{j}}{\sigma_{j}}\right )^{2}\right) -  \nonumber \\
& \gamma \left(\frac{n+1}{2},\left(\frac{\left(p_{i}-\mu_{j} \right)}{\sigma_{j}}\right)^{2} \right)\Bigg ) \Bigg \} \nonumber ,\textrm{  and} \nonumber \\
& \gamma \left (\centerdot,\centerdot \right) \textrm{is the lower incomplete gamma function }  \nonumber 
\end{align}

\subsubsection{The conditional density function} \label{sec323}
The conditional density function of $S_{i}$ where it is known that $P_{i}=p_{i}$ is
\begin{align*} %\label{eq337}
&f_{S_{i}\mid P_{i}}\left(s_{i} \mid p_{i}\right) \nonumber \\
&=\frac{\sqrt{2} p_{i}^{1-a_{j}u_{j}}} {C_{7,j}\sigma_{j}} \frac{ s_{i}^{a_{j}u_{j}-1}\left(1- \left(1-c_{j}\right) \left(\frac{s_{i}}{d_{j}}\right)^{a_{j}}\right)^{v_{j}-1}e^{-\frac{\left(p_{i}-s_{i}-\mu_{j}\right)^{2}}{2\sigma_{j}^{2}}}} { \left(1+c_{j}\left(\frac{s_{i}}{d_{j}}\right)^{a_{j}} \right)^{u_{j}+v_{j}}}
\end{align*}

\subsubsection{The corrected background intensity} \label{sec324}
The corrected background intensity under this generalized beta-normal convolution is
\begin{align}\label{eq338}
& p_{i}\frac{C_{8,j}}{C_{7,j}}
\end{align}
where
\begin{align*} 
C_{8,j} =& \sum \limits_{l=0}^{\infty} \sum_{m=0}^{\infty} \sum_{n=0}^{\infty}\left \{ \frac{ (-1)^{l+m} \left(1-c_{j} \right)^{l} c_{j}^{m} \dbinom{v_{j}-1}{l} \dbinom{u_{j}+v_{j}+m-1}{m}   } {d_{j}^{a_{j}(l+m)} \left(p_{i}-\mu_{j}\right)^{n}}  \times \right. \nonumber \\
& \left. \dbinom{a_{j}\left(u_{j}+l+m\right)}{n} \left(p_{i}-\mu_{j}\right)^{a_{j}(l+m)}  \sigma_{j}^{n} 2^{\frac{n}{2}} \Bigg (\gamma \left ( \frac{n+1}{2} , \left (\frac{\mu_{j}}{\sigma_{j}} \right )^{2}\right )-   \right. \nonumber \\
& \left. \gamma \left (\frac{n+1}{2} , \left (\frac{\left(p_{i}-\mu_{j}\right)}{\sigma_{j}} \right )^{2} \right ) \Bigg )\right \}  \nonumber , \textrm{  and} \nonumber \\
& \gamma \left (\centerdot,\centerdot \right) \textrm{  is the lower incomplete gamma function  } \nonumber 
\end{align*}

\subsubsection{The likelihood function} \label{sec325}
The likelihood function (\textbf{L}) to estimate  $ a_{j}, c_{j}, d_{j}, u_{j}, v_{j},  \mu_{j}$ and $\sigma_{j}^{2}$ is
\begin{align*} %\label{eq339}
=&\prod \limits_{i=1}^{I}\frac{\abs{a_{j}} p_{i}^{a_{j}u_{j}-1} C_{7,j}} {2\sqrt{\pi} d_{j}^{a_{j}u_{j}} \textrm{B}\left(u_{j},v_{j} \right) } \prod \limits_{w=1}^{W} \frac{e^{-\frac{1}{2\sigma_{j}^{2}} \left(b_{0w}-\mu_{j} \right)^{2}}} {\sqrt{2\pi} \sigma_{j}}
\end{align*}

The log-likelihood function $l$ is
\begin{align} \label{eq340}
=&\sum \limits_{i=1}^{I} \left \{\ln \left(\abs{a_{j}} \right)+\left(a_{j}u_{j} -1\right) \ln \left(p_{i}\right)+\ln (C_{7,j}) -\ln(2) - \frac{1}{2} \ln(\pi) -a_{j}u_{j} \ln \left(d_{j}\right) -  \right. \nonumber  \\
& \left. \ln \left(\textrm{B}\left(u_{j},v_{j}\right)\right) \right \} +  \sum \limits_{w=1}^{W} \left \{ -\frac{\left(b_{0w}-\mu_{j}\right)^{2}}{2\sigma_{j}^{2}} - \frac{1}{2} \left(\ln (2) +\ln (\pi) \right) -\ln (\sigma_{j}) \right \}
\end{align}
%\color{blue}
The likelihood equations are as follows
\begin{align*} %\label{eq341}
\frac{\partial l}{\partial \mu_{j}} = &  \sum \limits_{w=1}^{W} \left( \frac{\left(b_{0w} - \mu_{j}\right)}{\sigma_{j}^{2}} \right ) = 0
\end{align*}
\begin{align*} %\label{eq342}
\frac{\partial l}{\partial \sigma_{j}} = &  \sum \limits_{w=1}^{W} \left( \frac{\left(b_{0w} - \mu_{j}\right)^{2}}{\sigma_{j}^{3}} - \frac{1}{\sigma_{j}} \right ) = 0
\end{align*}
\begin{align*} %\label{eq343}
\frac{\partial l}{\partial a_{j}} = &  \sum \limits_{i=1}^{I} \left( \frac{1}{\abs{a_{j}}} + u_{j} \ln \left( p_{i} \right) + \frac{\frac{\partial C_{7,j}}{\partial a_{j}}}{C_{7,j}} - \mu_{j} \ln \left( d_{j} \right ) \right )=0
\end{align*}
\begin{align*} %\label{eq344}
\frac{\partial l}{\partial c_{j}} = &  \sum \limits_{i=1}^{I} \left( \frac{\frac{\partial C_{7,j}}{\partial c_{j}}}{C_{7,j}}  \right ) = 0
\end{align*}
\begin{align*} %\label{eq345}
\frac{\partial l}{\partial d_{j}} = &  \sum \limits_{i=1}^{I} \left( \frac{\frac{\partial C_{7,j}}{\partial d_{j}}}{C_{7,j}} - \frac{a_{j}\mu_{j}}{d_{j}} \right ) = 0
\end{align*}
\begin{align*} %\label{eq346}
\frac{\partial l}{\partial u_{j}} = &  \sum \limits_{i=1}^{I} \left( a_{j} \ln \left(p_{i} \right) + \frac{\frac{\partial C_{7,j}}{\partial u_{j}}}{C_{7,j}} - \frac{\frac{\partial \textrm{B} \left(u_{j}, v_{j} \right) }{\partial u_{j}}}{\textrm{B} \left(u_{j}, v_{j} \right) } \right ) = 0
\end{align*}
\begin{align*} %\label{eq347}
\frac{\partial l}{\partial v_{j}} = & \sum \limits_{i=1}^{I} \left( \frac{\frac{\partial C_{7,j}}{\partial v_{j}}}{C_{7,j}} - \frac{\frac{\partial \textrm{B} \left(u_{j}, v_{j} \right) }{\partial v_{j}}}{\textrm{B} \left(u_{j}, v_{j} \right) } \right ) = 0
\end{align*}
where
\begin{align*} %\label{eq348}
\frac{\partial \textrm{B} \left(u_{j}, v_{j} \right) }{\partial u_{j}} &= \Gamma \left(v_{j} \right ) \frac{\Gamma\left(u_{j}\right) \left(-\gamma + \sum \limits_{k=1}^{u_{j}-1} \frac{1}{k} \right) - \Gamma\ \left(u_{j} \right) \left( -\gamma + \sum \limits_{k=1}^{u_{j}+v_{j}-1} \frac{1}{k}\right)}{\Gamma \left( u_{j}+v_{j} \right)} \nonumber \\
&=  \textrm{B} \left(u_{j}, v_{j} \right) \left( \sum \limits_{k=1}^{u_{j}-1} \frac{1}{k} - \sum \limits_{k=1}^{u_{j}+v_{j}-1} \frac{1}{k} \right)
\end{align*}
\begin{align*} %\label{eq349}
\frac{\partial \textrm{B} \left(u_{j}, v_{j} \right) }{\partial v_{j}}&= \Gamma \left(u_{j} \right ) \frac{\Gamma\left(v_{j}\right) \left(-\gamma + \sum \limits_{k=1}^{v_{j}-1} \frac{1}{k} \right) - \Gamma\ \left(v_{j} \right) \left( -\gamma + \sum \limits_{k=1}^{u_{j}+v_{j}-1} \frac{1}{k}\right)}{\Gamma \left( u_{j}+v_{j} \right)} \nonumber \\
&= \textrm{B} \left(u_{j}, v_{j} \right) \left( \sum \limits_{k=1}^{v_{j}-1} \frac{1}{k} - \sum \limits_{k=1}^{u_{j}+v_{j}-1} \frac{1}{k} \right)
\end{align*}
and suppose $C_{7,j}$ is written as $\sum \limits_{l=0}^{\infty} \sum \limits_{m=0}^{\infty} \sum \limits_{n=0}^{\infty} C_{7lmn}$ then
\begin{align*} %\label{eq350}
\frac{\partial C_{7,j}}{\partial a_{j}}=& \sum \limits_{l=0}^{\infty} \sum \limits_{m=0}^{\infty} \sum \limits_{n=0}^{\infty} \left [ C_{7lmn} \Bigg ( \left( l+m \right) \ln \left ( \frac{p_{i} - \mu_{j}}{d_{j}}\right) + \left( u_{j}+l+m \right) \times \right. \nonumber \\
& \left. \left ( \sum \limits_{k=1}^{a_{j} \left( u_{j}+l+m\right) -1} \frac{1}{k}- \sum \limits_{k=1}^{a_{j} \left( u_{j}+l+m\right)-n -1} \frac{1}{k} \right) \Bigg ) \right ]
\end{align*}
\begin{align*} %\label{eq351}
&\frac{\partial C_{7,j}}{\partial c_{j}} = \sum \limits_{l=0}^{\infty} \sum \limits_{m=0}^{\infty} \sum \limits_{n=0}^{\infty} \left [ C_{7lmn} \left (\frac{\left(1 - c_{j} \right)m-lc_{j}}{c_{j}\left(1 - c_{j} \right)} \right) \right ]
\end{align*}
\begin{align*} %\label{eq352}
&\frac{\partial C_{7,j}}{\partial d_{j}}  = \sum \limits_{l=0}^{\infty} \sum \limits_{m=0}^{\infty} \sum \limits_{n=0}^{\infty} \left [ C_{7lmn} \left (-\frac{a_{j} \left(l+m\right)}{d_{j}} \right ) \right ]
\end{align*}
\begin{align*} %\label{eq353}
\frac{\partial C_{7,j}}{\partial u_{j}} =& \sum \limits_{l=0}^{\infty} \sum \limits_{m=0}^{\infty} \sum \limits_{n=0}^{\infty} \left [ C_{7lmn} \Bigg ( \left ( \sum \limits_{k=1}^{u_{j}+v_{j}+m-1} \frac{1}{k}- \sum \limits_{k=1}^{u_{j}+v_{j}-1} \frac{1}{k} \right) + \right. \nonumber \\
& \left. a_{j} \left ( \sum \limits_{k=1}^{a_{j} \left( u_{j}+l+m\right)-1} \frac{1}{k} - \sum \limits_{k=1}^{a_{j} \left( u_{j}+l+m\right)-n-1} \frac{1}{k}\right) \Bigg ) \right ]
\end{align*}
\begin{align*} %\label{eq354}
\frac{\partial C_{7,j}}{\partial v_{j}} =& \sum \limits_{l=0}^{\infty} \sum \limits_{m=0}^{\infty} \sum \limits_{n=0}^{\infty} \left [ C_{7lmn} \Bigg ( \left ( \sum \limits_{k=1}^{v_{j}-1} \frac{1}{k}- \sum \limits_{k=1}^{v_{j}-l-1} \frac{1}{k} \right) + \right. \nonumber \\
& \left. a_{j} \left ( \sum \limits_{k=1}^{u_{j}+v_{j}+m-1} \frac{1}{k} - \sum \limits_{k=1}^{ u_{j}+l+m-n-1} \frac{1}{k}\right) \Bigg ) \right ]
\end{align*}
%\color{black}
\vspace{1.5cc}
\section{Discussion and remarks} \label{sec4}
We have studied the additive models of background correction for BeadArrays and proposed the generalized model where the true intensity and the noise are assumed to be skewed distribution and where the true intensity is a skewed but the noise is symmetric distribution. In this paper, we have shown the corrected background intensity value of the proposed models.

This proposed model is a generalization of the available convolution models as in papers \cite{Bol03}, \cite{Iri03a,Iri03b,Iri06}, \cite{Che11}, \cite{Faj14}, \cite{Pla12} and \cite{Xie09}. The generalization comes from the property of the tree-generalized beta distributions \cite{Don95} and is explained in \cite{Don84} and \cite{Don95}. The parameters of the generalized beta distribution are $ a, d, c, u$ and $v$. The gamma, exponential and lognormal distributions are special cases of the generalized beta distribution. 

The gamma distribution is the generalized beta distribution when $c=1, v \rightarrow \infty, d=\beta v^{\frac{1}{a}}$ and $a=1$; the exponential distribution is the generalized beta distribution when $c=1, v \rightarrow \infty, d=\beta v^{\frac{1}{a}}$ and $a=1, p=1$;  and the lognormal distribution is the generalized beta distribution when $c=1,v \rightarrow \infty, d=\beta v^{\frac{1}{a}}$ and $\beta=(\sigma^{2}a^{2})^{\frac{1}{2}}, u=\frac{(a\mu+1)}{\sigma^{2}a^{2}}$ and $a \rightarrow 0$.

There are some aspects to be considered while implementing these models:
\begin{enumerate}
\item{parameters estimation } \\
In parameters estimation, there are some methods have been suggested by some researchers. Mc Donald and Xu \cite{Don95} used and suggested: the method of maximum likelihood (also was used by Fajriyah \cite{Faj05a,Faj05b,Faj08}), the method of moments and the maximum product spacing estimation.

When $c=1$, the generalized beta distribution is a generalized beta of the second kind. Graf and Nedyalkova \cite{Gra10} and Graf et al. \cite{Gra11} have observed that the pseudo maximum likelihood (Huber \cite{Hub67}, Freedman \cite{Fre06} and  Pfeffermann et al. \cite{Pfe98}), the nonlinear least squares on the quantile function (Dagum \cite{Dag77}) and the nonlinear fit for indicator  can be implemented to estimate the parameters of the generalized beta of the second kind. The available VGAM \emph{package} in R helps to estimate the parameters of this distribution.

The existing convolution models use various methods:
\begin{enumerate}
\item{the \emph{ad-hoc} method which is implemented by the RMA method, more details can be found in \cite{Iri03a, Iri03b, Iri06}, \cite{Gee06} and \cite{Xie09}}
\item{Markov chain Monte Carlo simulations, more details can be found in \cite{Din08}}
\item{Maximum likelihood, nonparametrics and method of moments, more details can be found in \cite{Che11}, \cite{Faj14} and \cite{Xie09}}
\item{Plug-in method, more details can be found in \cite{Faj14}}
\item{Fast Fourier transform, more details can be found in \cite{Pla12}}
\end{enumerate}
In general, we first need to provide the initial parameters to optimize the log-likelihood function in Equations (\ref{eq310}) and (\ref{eq340}). The initial parameters of the noise are easily provided since the benchmarking data set of the negative control probes is available publicly. The initial parameters of the true intensity can be estimated from the observed intensity data substracted by the mean (or median) of the negative control intensity. 

Secondly, once the initial parameters are available, then they will be used to optimize the likelihood function by implementing the optimization method. There are some packages in $R$ which can be used to compute the parameters of the model, for example the \emph{optim} or \emph{optimx} package. These parameters are then used to compute the corrected background intensity based on the formula of the choosen model. Remember that the background correction is implemented for each array.

\item{the corrected background intensity computation} \\
The corrected background intensity computation includes computations of the infinite summations: $C_{5,j}, C_{6,j}, C_{7,j}$ and $C_{8,j}$. In the author's experience (in \cite{Faj14}) these infinite summations are close to being constant after certain terms. As a consequence, the ratios of $\frac{C_{6,j}}{C_{5,j}}$ and $\frac{C_{8,j}}{C_{7,j}}$ are able to be computed. Therefore the difficulty in computing the summations used to compute the corrected background intensity can be eliminated. A sophisticated program written in $R$, $C$, $Python$ and its paralellisation, could help to speed up the computation.

\item{the benchmarking data set} \\
During the implementation of this generalized estimator, the Illumina users need to be aware of the availability of the Illumina Spike-in data set. Once the model is fitted into this data set, the model can then be used to adjust the intensity value. 

Apart from the benchmarking criteria for the Affymetrix GeneChips, in the author's knowledge, the benchmarking criteria for the Illumina BeadArrays have not been formalized yet. Some researchers, i.e. \cite{Che11}, \cite{Pla12}, \cite{Sha06} and \cite{Xie09} have developed the criteria to assess which background correction methods perform better than the others for the Illumina BeadArrays. 

These criteria together with the criteria in the Affycomp \textit{package} (\cite{Cop04} and \cite{Iri13}) can be used as the benchmarking criteria for the Illumina BeadArrays. These have been implemented by Fajriyah \cite{Faj14}. The method which has been used by Shi et al. \cite{Shi10} also can be used to assess the best performance of the background correction methods.

%In case there is no benchmarking data set, the methods which have been implemented by \cite{Che11},  \cite{Faj14a}, \cite{Pla11,Pla12}, and \cite{Xie09} can be used to assess the best convolution model for the background correction.

%Beside the benchmarking criteria, it is possible to assess which background correction method actually performs the best in the pre-processing of the intensity of the Illumina BeadArrays, based on the precision and bias criteria. The study of Shi et al. \cite{Shi10}, based on the four data sets, show that the normal-exponential convolution of background correction using the control probes gives the least biased foldchanges of the given false discovery rate. 

\item{the negative control data set} \\
It is possible that the negative control probes set data is unavailable. In this case, we can adapt the proposed model to the convolution model for background correction without the negative control probes intensities, as in the RMA model. %and \emph{normexp} \cite{Sil09} of the \emph{limma} packages
\end{enumerate}

The application of this generalized model towards other platforms, such as the Affymetrix, is possible by considering the points above.

%In our future research we would like to address the implementation of the proposed model by considering the parameter estimation methods and the best fitted model in the benchmarking and other criteria such as the precision and bias at Shi et al. \cite{Shi10} paper, at both of the benchmarking and real data sets.

\vspace{1.5cc}
\textbf{Acknowledgements:} \\
This paper is part of the author's PhD dissertation written under the direction of Professor Istv\'an Berkes. We would like to thank \textit{Paulo Canas Rodrigues, PhD} for his comments. Financial support from the Austrian Science Fund (FWF), Project P24302-N18 is gratefully acknowledged. We would also like to thank the anonymous reviewers for their valuable remarks in leading to an improvement of this paper.

\textbf{Conflicts of interest:} None

%\bibliographystyle{pccp} 
%\bibliographystyle{jthcarsu} 
%\bibliographystyle{apa} 
%\bibliographystyle{achemso} 
%\bibliography{emaRef}

\begin{thebibliography}{100}
\bibitem{Hub04} Huber, W., von Heydebreck, A. and Vingron,M.,  Error models for microarray intensities, Technical Report Paper 6, Bioconductor Project Working Papers, 2004.
\bibitem{Hub05a} Huber, W., von Heydebreck, A. and Vingron,M., An introduction to low-level analysis methods of DNA microarray data Technical Report Paper 9, Biocon- ductor Project Working Papers, 2005a.
\bibitem{Hub05b} Huber, W., Irizarry, R. A.  and Gentleman, R. , Bioinformatics and Computational Biology Solutions Using R and Bioconductor; chapter Prepocessing Overview, Springer, 2005b.
\bibitem{Bol03} Bolstad, B. M., Irizarry, R. A., Astrand, M. and Speed, T. P., A Comparison of Normalization Methods for High Density Oligonucleotide Array Data Based on Bias and Variance, Bioinformatics, 2003; 19(2): 185-193.
\bibitem{Iri03a} Irizarry, R. A., Bolstad, B. M., Collin, F., Cope, L. M., Hobbs, B.  and Speed, T. P., Summaries of Afymetrix GeneChip probe level data, Nucleic Acids Research, 2003a; 31(4):e15. \url{doi: 10.1093/nar/gng015}
\bibitem{Iri03b} Irizarry, R. A., Hobbs, B., Collin, F., Beazer-Barclay, Y. D. ,  Antonellis, K. J., Scherf, U. and Speed, T. P., Exploration, Normalization and Summaries of High Density Oligonucleotide Array Probe Level Data, Biostatistics, 2003b; 4(2): 249-264.
\bibitem{Iri06} Irizarry, R. A., Wu, Z., and Jaffee, H. A., Comparison of Affymetrix geneChip expression measures, Bioinformatics, 2006; 22(7): 789-794.
\bibitem{LiW01} Li, C., and Wong, W. H., Model-based analysis of oligonucleotide arrays: Expression index computation and outlier detection, Proceeding National Academy of Sciences, 2001; 98(1): 31-36.
\bibitem{Sil09} Silver, J. D., Ritchie, M. E., and Smyth, G. K., Microarray background correction: maximum likelihood estimation for the normal-exponential convolution model, Biostatistics, 2009; 10: 352-363.
\bibitem{Wu04} Wu, Z., Irizarry,R. A., Gentleman, R., Martinez-Murillo, F. and Spencer, F., A model-based background adjustment for oligonucleotide expression arrays, Journal of the American Statistical Association, 2004; 99(468): 909-917.
\bibitem{Din08} Ding, L.-H., Xie, Y., Park, S., Xiao, G. and Story, M. D., Enhanced identification and biological validation of differential gene expression via Illumina whole genome expression arrays through the use of the model-based background correction methodology, Nucleic Acids Research, 2008; 36(10: e58).
\bibitem{Xie09} Xie, Y., Wang, X. and Story, M. D., Statistical methods of background correction for Illumina BeadArray data, Bioinformatics, 2009; 25(6): 751-757.
\bibitem{Che11} Chen, M., Xie, Y. and Story, M. D., An Exponential-Gamma Convolution Model for Background Correction of Illumina BeadArray Data, Communication in Statistics: Theory and Methods, 2011; 40(17): 3055-3069.
\bibitem{Pla12} Plancade, S., Rozenholc, Y. and Lund, E., Generalization of the normal- exponential model: exploration of a more accurate parameterisation for the signal distribution on Illumina BeadArrays, BMC Bioinformatics, 2012; 13(329).
\bibitem{Faj14} Fajriyah, R., A Study of convolution models for background correction of BeadArrays, accepted paper at Austrian Journal of Statistics, 2014.
\bibitem{Pos11} Posekany, A., Felsenstein, K. and Sykacek, P. Biological assessment of robust noise models in microarray data analysis, Bioinformatics, 2011; 27(6): 807-814.
\bibitem{Don95} McDonald, J. B. and Xu, Y. J., A generalization of the beta distribution with applications, Journal of Econometrics, 1995; 66: 133-152.
\bibitem{Lee08} Leemis, L. M. and McQueston, J. T., Univariate Distribution Relationships, The American Statistician, 2008; 62(1): 45-53.
\bibitem{Don84} McDonald, J. B.,Some generalized functions for the distribution of income, Econometrica, 1984; 52(3):647-663.
\bibitem{Faj05a} Fajriyah, R., Statistical analysis of the economic performance in Indonesia, Part I - Simplex method, 55th ISI Session Conference, 2005a.
\bibitem{Faj05b} Fajriyah, R., Statistical analysis of the economic performance in Indonesia, Part II - Grad method, ICREM 2 Conference, INSPEM, University Putra Malaysia, 2005b.
\bibitem{Faj08} Fajriyah, R., The pdfs estimation by grad method and its Gini index, Karya Asli Lorekan Matematik, 2008; 1(2): 021-027.
\bibitem{Gra10} Graf, M. and Nedyalkova, D., Fitting the Generalized Beta Distribution of the Second Kind to the Empirical Income Distribution from the Aggregate Laeken Indicators, 2010. Available from URL \url{http://www.statistik.tuwien.ac.at/ameli/presentations/Fri1/GrafNedyalkova1.pdf.pdf}  [accessed March 3, 2014]
\bibitem{Gra11} Graf, M., Nedyalkova, D., M\"{u}nnich,  R., Seger, J. and Zins, S., Parametric Estimation of Income Distributions and Indicators of Poverty and Social Exclusion, Technical Report 2.1, AMELI, 2011.
\bibitem{Hub67} Huber, P. J., In The behavior of maximum likelihood estimates under nonstandard conditions, Proceedings of the Fifth Berkeley Symposium on Mathematical Statistics and Probability, Vol. 1: Statistics, 221-233, Berkeley ,California, Univ. Calif. Press, 1967. 
\bibitem{Fre06}  Freedman, D. A., On the so-called "Huber sandwich estimator" and "robust standard errors", The American Statistician, 2006; 60: 299-302.
\bibitem{Pfe98}  Pfeffermann, D., Skinner, C. J., Holmes, D. J., Goldstein, H., and Rasbash, J., Weighting for unequal selection probabilities in multilevel models, Journal of the Royal Statistical Society B, 1998; 60(Part 1): 23-40.
\bibitem{Dag77} Dagum, C., A New Model of Personal Income Distribution: Specification and Estimation, Economie Appliquée, 1977; 30: 413-437.
\bibitem{Gee06} McGee, M. and Chen, Z., Parameter estimation for the convolution model for background correction of affymetrix genechip data, Statistical Applications in Genetics and Molecular Biology, 2006; 5(24). \url{doi:10.2202/1544-6115.1237}.
\bibitem{Sha06} Shamilov, A., Kantar, Y. M. and Usta, I., In On a Functional defined by means of Kullback-Leibler Measure and Its Statistical Applications, Proceedings of the 9th WSEAS International Conference on Applied Mathematics, 632-637, 2006.
\bibitem{Cop04} Cope, L. M., Irizarry, R. A., Jaffee, H. A., Wu, Z. and Speed, T. P. , A benchmark for Affymetrix GeneChip expression measures, Bioinformatics, 2004; 20: 323-331.
\bibitem{Iri13} Irizarry, R. A., and Wu, Z., affycomp: Graphics Toolbox for Assessment of Affymetrix Expression Measures. R package version 1.38.0 (with contributions from Simon Cawley) ed., 2013.
\bibitem{Shi10} Shi, W., Oshlack, A. and Smyth, G. K., Optimizing the noise versus bias trade-off for Illumina whole genome expression Beadchips, Nucleic Acids Research, 2010; 38(22): e204. \url{doi: 10.1093/nar/gkq871}.
\end{thebibliography}
%\begin{appendices}

\end{document}